D.N. Makovetskii

# Optical-Wavelength Paramagnetic Phaser (Lecture Notes).
## Section 3.1. Nonlinear Balance Equations of Motion

In this Section we present a detailed description of the simplest nonlinear model for an optical wavelength paramagnetic phaser[*], which is an acoustic analog of the class-B lasers. Despite of its simplicity, this model gives a satisfactory explanation of experimental data for optical-wavelength paramagnetic phasers based on high-quality acoustic Fabry-Perot resonators. In particular, this model was successfully used both for qualitative and quantitative interpretation of deterministic chaotic motions observed in spin-phonon system of a nonautonomous ruby phasers at liquid helium temperatures (see arXiv:0704.0123v1 [nlin.CD]).

**3.1.1. Equations of motion for the balance model of spin-phonon interaction**

Let us consider the spin-system of impurity paramagnet, having three enegy levels $E_3 > E_2 > E_1$, where the transition of signal ($S$-transition) $E_1 \leftrightarrow E_2$ is enabled for interaction with coherent optical-wavelength field (microwave ultrasound called also hypersound) of the specified direction and polarization, and the transition of pumping ($P$-transition) $E_1 \leftrightarrow E_3$ is enabled for magnetic dipole interaction with appropriate microwave electromagnetic pumping field. The third transition ($F$-transition) is idler, and by definition its frequency $\Omega_F \equiv (E_3 - E_2)/\hbar$ is not equal or multiple of $S$-transition frequency $\Omega_S \equiv (E_2 - E_1)/\hbar$. Besides of this, $\Omega_F$ is not integer-valued divider of $P$-transition frequency $\Omega_P \equiv (E_3 - E_1)/\hbar$.

In doped paramagnets of the A-type (where ground orbital state is singletic) the longitudinal relaxation times $T_{1S}, T_{1P}, T_{1F}$ for all the pointed transition at low temperatures are of many orders smaller than transverse relaxation times $T_{2S}, T_{2P}, T_{2F}$. Therefore, without loss of generality, one may choose the amplitude of microwave electromagnetic pumping field $H_{1P}$ such that the populations $n_1$ and $n_3$ of the spin-levels $E_1$ and $E_3$ would be already equal, but the broadening of these levels still remains the same as at $H_{1P} = 0$ (with the accuracy up to higher orders). In the other words, we shall assume that the next inequalities take place:

$$Z_P \gg 1; \quad (T_{2P}/T_{1P})Z_P \ll 1, \tag{3.1.1}$$

or, in slightly other form:

$$Y_P T_{1P} \gg 1 \gg Y_P T_{2P}, \tag{3.1.2}$$

where $Z_P$ – the $P$-transition saturation factor:

$$Z_P = \frac{1}{4} T_{1P} T_{2P} \gamma_P^2 H_{1P}^2, \tag{3.1.3}$$

and $Y_P$ – the probability of interaction of the pumping field with the $P$-transition. Here $\gamma_P$ – effective hyromagnetic ratio for this spin transition (accounting direction and polarization of the vector $H_{1P}$).

---
[*] Phaser = Phonon Amplification by Stimulated Emission of Radiation

Moreover, the phonon lifetime $T_{CAV}$ in a high-quality optical-wavelength acoustic resonator, e.g. acoustic Fabry-Perot resonator (AFPR), usually meets a condition [1 - 7]:

$$T_{1S} \gg T_{CAV} \gg T_{2S}, \tag{3.1.4}$$

and for the microwave electromagnetic pumping resonator the similar inequalities take place for the photon lifetime. And at last it is supposed that pump resonator has not eigen frequencies in the vicinity of $\Omega_F$.

Under these circumstances, it is valid to use the equations of the balance approximation [8] for the calculation of the $S$-transition population inversion $\Delta N \equiv n_2 - n_1$, which (with accounting of the condition $n_3 = n_1$) reduces to the form:

$$\left.\begin{array}{l}\dfrac{\mathrm{d}n_2}{\mathrm{d}t}=-(W_{21}+W_{23})n_2+W_{12}n_1+W_{32}n_1-(n_2-n_1)Y_\mathrm{S};\\[2mm]-\dfrac{\mathrm{d}n_1}{\mathrm{d}t}=(W_{12}+W_{13})n_1-W_{21}n_2-W_{31}n_1-(n_2-n_1)Y_\mathrm{S}\end{array}\right\},\qquad(3.1.5)$$

where $Y_\mathrm{S}$ – the probability of interaction of hypersonic field with the spin's $S$-transition; $W_{ij}$ – the probabilities of the longitudinal spin relaxation [8, 9]. Following the works [1 - 7] we have

$$Y_\mathrm{S}=T_{2\mathrm{S}}k_u^2 U^2 |F_{12}|^2/\hbar^2 \equiv Z_\mathrm{S}/T_{1\mathrm{S}},\qquad(3.1.6)$$

where $k_u=\Omega_\mathrm{S}/V_u$; $U$ and $V_u$ –accordingly the amplitude and the phase velocity of hypersound; $F_{12}$ – the factor of spin-phonon interaction for the $S$-transition at the specified values of hypersound direction and polarization. The general expressions for $F_{12}$ one can find in the work [10]. For the case of longitudinal hypersound propagating along the crystallographic axis $Oz$ of the third or higher order (as in the 9 GHz ruby phaser [1 - 7]) we have:

$$F_{12}=\dfrac{\partial}{\partial\varepsilon_{zz}}\langle\psi_1|\hat{\mathrm{H}}_{su}|\psi_2\rangle.\qquad(3.1.7)$$

Here $\varepsilon_{zz}$ – the component of elastic deformations tensor; $\psi_1$ and $\psi_2$ – the wave functions belonging to the spin-levels $E_1$ and $E_2$; $\hat{\mathrm{H}}_{su}$ – Hamiltonian of spin-phonon interaction.

Using the approximation $\hbar\Omega_\mathrm{P}\ll 3k_\mathrm{B}\theta$ (where $k_\mathrm{B}$ – Boltzmann constant; $\theta$ – thermostate temperature), from (3.1.5) we receive

$$\dfrac{\mathrm{d}(\Delta N)}{\mathrm{d}t}=-2(\Delta N)Y_\mathrm{S}+\dfrac{(\Delta N)_0-(\Delta N)}{T_{1\mathrm{S}}^{(P)}},\qquad(3.1.8)$$

where

$$(\Delta N)_0=(-4s_\theta W_\mathrm{S}+2f_\theta W_\mathrm{F}-2p_\theta W_\mathrm{P})W_\mathrm{E}^{-1}N_c;$$

$$T_{1S}^{(P)} = 3/W_E; \qquad N_c = n_1 + n_2 + n_3;$$

$$s_\theta = \hbar\Omega_S/3k_B\theta; \quad W_S = \frac{W_{12} + W_{21}}{2};$$

$$p_\theta = \hbar\Omega_P/3k_B\theta; \quad W_P = \frac{W_{13} + W_{31}}{2};$$

$$f_\theta = \hbar\Omega_F/3k_B\theta; \quad W_P = \frac{W_{23} + W_{32}}{2};$$

$$W_E = (6 + 2s_\theta)W_S + (3 - f_\theta)W_F + 2p_\theta W_P;$$

The effective longitudinal relaxation time (atomic relaxation time of phaser's signal channel) $T_{1S}^{(P)}$ that figures in (3.1.8) is not the usual spin-lattice relaxation time $T_{1S}^{(0)}$ used in investigations of passive systems, because pumping (hidden in (3.1.5) due to $n_1 = n_3$) leads to renormalization of longitudinal relaxation time. For example, at $W_F \gg W_S, W_P$ we find $T_{1S}^{(P)} \approx 1/W_F \ll T_{1S}^{(0)}$. From now on we shall omit the superscript in $T_{1S}^{(P)}$.

Let us introduce the dynamical variable $\overline{M}$, which is proportional to the average intensity of the phonon stimulated emission (SE) in a phaser:

$$\overline{M} = 2\rho'\Omega_S(\overline{U})^2\hbar^{-1} \equiv 2\overline{Y}_S/B', \qquad (3.1.9)$$

where $B' = T_{2S}\Omega_S|F_{12}|^2/\rho' V_u^2\hbar$; $\rho'$ – crystal density, and the upper bar means the averaging through the AFPR volume.

On the basis of the wave equation for hypersound in active paramagnetic medium [10] and using the approximations [11], we receive equation for the first derivative of the phonon intensity in the phaser:

$$\frac{d\overline{M}}{dt} = B'\overline{M}\,\overline{N} - \frac{\overline{M}}{T_{1S}}, \qquad (3.1.10)$$

where $\overline{N} = \overline{(\Delta N)}$.

Averaging also (3.1.8), we find

$$\frac{d\overline{N}}{dt} = -B'\overline{M}\,\overline{N} + \frac{\overline{N}_0 - \overline{N}}{T_{1S}}. \tag{3.1.11}$$

The system (3.1.10) – (3.1.11) represents the simplest nonlinear equations of motion for spin-phonon system in three-level autonomous phaser. This system is isomorphic to equations of motion for two-level autonomous class-B laser [12] – owing to pump equations reduction. Introducing $\tau = t/T_{1S}$ we proceed with dimensionless form of these equations. To the effect we shall use the following dynamical variables:

$$J(\tau) = B'T_{1S}\overline{M} \equiv 2\overline{Z}_S; \quad n(\tau) = \overline{N}/\overline{N}_{tr}, \tag{3.1.12}$$

and the next control parameters:

$$A = \overline{N}_0/\overline{N}_{tr}; \quad B = T_{1S}/T_{CAV}, \tag{3.1.13}$$

where $\overline{N}_{tr}$ – the threshold value of inverted spin-level population difference correspondent to phonon SE self-excitation:

$$\overline{N}_{tr} = \frac{1}{B'T_{CAV}} = \frac{\rho' V_u^2 \hbar}{T_{CAV} T_{2S} \Omega_S |F_{12}|^2}. \tag{3.1.13}$$

Now the system (3.1.10), (3.1.11) reshapes to:

$$\left.\begin{array}{l} \dfrac{dJ}{d\tau} = BJ(n-1) \\ \dfrac{dn}{d\tau} = A - n(J+1) \end{array}\right\}, \tag{3.1.14}$$

and represents the autonomous flow system [13] in a two-dimensional phase space:

$$\left.\begin{array}{l} \dfrac{dz_1}{d\tau} = \Phi_1(z_1, z_2; c_1, c_2); \\ \dfrac{dz_2}{d\tau} = \Phi_2(z_1, z_2; c_1, c_2) \end{array}\right\}, \tag{3.1.15}$$

where $z_{1,2}$ – dynamical variables, $z_1 = J$; $z_2 = n$; $\Phi_{1,2}$ – nonlinear functions of the dynamical variables, dependent on control parameter set ($c_1 = A$; $c_2 = B$) as well:

$$\left.\begin{aligned}\Phi_1 &= Bz_1(z_2 - 1); \\ \Phi_2 &= A - z_2(z_1 + 1)\end{aligned}\right\}. \tag{3.1.16}$$

### 3.1.2. Relaxation frequency of an autonomous phaser

Lyapunov stability of the special points – stationary solutions of the system (3.1.15), and the type of these special poins are defined by the equation:

$$\left|\left(\frac{\partial \Phi_i}{\partial z_j}\right)^{[st]} - \delta_{ij}\Lambda\right| = 0, \tag{3.1.17}$$

where $\Lambda = \Lambda' + i\Lambda''$ – dimensionless Lyapunov exponent (LE) spectrum; $\delta_{ij}$ – the Kroneker symbol, and the superscript $[st]$ means that derivative values have been taken at the stationary state points of phase space.

There are two stationary states $(d/d\tau = 0)$ of the system (3.1.15) with righ-hand parts (3.1.16):

$$\boxed{z_1^{[st1]} = 0; \quad z_2^{[st1]} = A}, \tag{3.1.18}$$

and

$$\boxed{z_1^{[st2]} = A - 1; \quad z_2^{[st2]} = 1}, \tag{3.1.19}$$

The expressions for $\partial \Phi_i/\partial z_j$, as it follows from (3.1.16), have the form:

$$\begin{aligned}\frac{\partial \Phi_1}{\partial z_1} &= B(z_2 - 1); & \frac{\partial \Phi_1}{\partial z_2} &= Bz_1; \\ \frac{\partial \Phi_2}{\partial z_1} &= -z_2; & \frac{\partial \Phi_2}{\partial z_2} &= -(z_1 + 1).\end{aligned} \tag{3.1.20}$$

From (3.1.17), (3.1.20) we receive equation for evaluation of our system's LE spectra in every special point:

$$\begin{vmatrix} B(z_2 - 1) - \Lambda & Bz_1 \\ -z_2 & -(z_1 + 1) - \Lambda \end{vmatrix} = 0. \qquad (3.1.21)$$

Expanding the determinant we get:

$$\Lambda^2 - [B(z_2 - 1) - z_1 - 1]\Lambda - \\ - B(z_2 - z_1 - 1) = 0. \qquad (3.1.22)$$

For the special point $[st1]$ from (3.1.18) and (3.1.22) we find:

$$\Lambda_{1,2}^{[st1]} = \frac{(A-1)B - 1}{2} \pm \\ \pm \sqrt{\left(\frac{(A-1)B - 1}{2}\right)^2 + (A-1)B} \qquad (3.1.23)$$

Thereby, the first special point $[st1]$ is the saddle:

$$\operatorname{Re} \Lambda_1^{[st1]} > 0; \quad \operatorname{Re} \Lambda_2^{[st1]} < 0; \\ \operatorname{Im} \Lambda_{1,2}^{[st1]} = 0. \qquad (3.1.24)$$

It means that all the trajectories in the phase space, excluding one lies strictly at the axis $J \equiv z_1 = 0$, are pushed from the point $[st1]$. Accordingly, whichever small disturbance of initial condition $J = 0$ takes away the imaging point from $[st1]$.

Next, for the second special point $[st2]$ we get:

$$\Lambda_{1,2}^{[st2]} = -\frac{A}{2} \pm \\ \pm \sqrt{\left(\frac{A}{2}\right)^2 - (A-1)B} \ . \qquad (3.1.25)$$

Taking in account $B \gg 1$ at $(4B)^{-1} \ll A-1 \ll 4B$ we have the following LE pair :

$$\Lambda_{1,2}^{[st2]} \approx -\frac{A}{2} \pm \sqrt{(1-A)B} . \qquad (3.1.26)$$

Then, by virtue of $A > 1$ there are:

$$\begin{aligned} \operatorname{Re}\Lambda_1^{[st2]} &= \operatorname{Re}\Lambda_2^{[st2]} < 0 ; \\ \operatorname{Im}\Lambda_{1,2}^{[st2]} &\neq 0. \end{aligned} \qquad (3.1.27)$$

Consequently $[st2]$ is the special point of the focus type, which incidentally is the single attractive set (attractor) of our system. Thus at the initial condition $J = J_0 > 0$ for the phonon SE supermode in the autonomous phaser the oscillating transient process takes place. The corresponding frequency has the name "relaxation frequency" and reads as the LE imaginary part modulo:

$$\widetilde{\omega}_{rel} = \left|\operatorname{Im}\Lambda^{[st2]}\right| = \sqrt{(A-1)B} , \qquad (3.1.28)$$

or, in the dimensioned form:

$$\omega_{rel} = \widetilde{\omega}_{rel}/T_{1S} = \sqrt{\frac{(A-1)}{T_{1S}T_{CAV}}} , \qquad (3.1.29)$$

### 3.1.3. Equations of motion for nonautonomous phaser

Small periodic perturbation $\omega_m \approx \omega_{rel}$ (modulation factor $k_m \ll 1$) of at least one control parameter in our system brings the destroying of the focus $[st2]$ resulting in soft appearance of the attractor $1\mathbf{P}_{B0}$ – limit cycle with exactly the same period as the external force period $T_m = 2\pi/\omega_m$ (the index "B0" in the limit cycle designation is the primary branch pointer). This fact is well known in laser dynamics (see e.g. [12]) and may be easy examined from the next equations of motions, where the control parameter $B$ is

modulated:

$$\frac{dJ}{d\tau} = BJ[n - 1 - k_m \cos(\widetilde{\omega}_m \tau)], \quad (3.1.30)$$

$$\frac{dn}{d\tau} = A - n(J + 1). \quad (3.1.31)$$

Here $\widetilde{\omega}_m = \omega_m T_{1S}$ – dimensionless frequency of modulation. The modulation of the control parameter $A$ gives qualitatively the same result if the depth of modulation stays small (see [14]).

With growing of the modulaton factor $k_m$ the behaviour of nonlinear oscillator, described by equations (3.1.30), (3.1.31), becomes very complicated [12, 14, 15], and only minor information one can get from purely analytical studies [16]. In Section 3.2 [17], a numerical integration of these equations will be presented. For the purpose of convenience let us transform the system (3.1.30) – (3.1.31) to another form and then convert it to the equivalent autonomous system.

Now we redefine our two control parameters introducing $\varepsilon_1$ and $\varepsilon_2$ instead of $A$ and $B$:

$$\left.\begin{array}{l} \varepsilon_1 = \dfrac{1}{\sqrt{(A-1)B}} \equiv (\widetilde{\omega}_{rel})^{-1} \equiv \dfrac{1}{\omega_{rel} T_{1S}}; \\[2mm] \varepsilon_2 = \sqrt{\dfrac{A-1}{B}} \equiv \dfrac{\widetilde{\omega}_{rel}}{B} \equiv \omega_{rel} T_{CAV} \end{array}\right\}, \quad (3.1.32)$$

or, for the inverse relationship:

$$A = 1 + \frac{\varepsilon_2}{\varepsilon_1}; \quad B = \frac{1}{\varepsilon_1 \varepsilon_2}. \quad (3.1.33)$$

As it is clear from (3.1.32), (3.1.29), at $A - 1 = O(1)$ we have $\varepsilon_1 \approx \varepsilon_2 \approx B^{-1/2} \ll 1$; $T_{rel} \equiv 2\pi/\omega_{rel} \approx \sqrt{T_{1S} T_{CAV}}$; and at small $A - 1 \approx 4\pi^2/B$ we receive $\varepsilon_1 \approx 1/2\pi$; $\varepsilon_2 \approx 2\pi/B$; $T_{rel} \approx T_{1S}$. The case $T_{rel} \approx T_{CAV}$ (as one may expect at $A \gg 1$) is unphysical because for large $A$ it is the pointed earlier restriction $A \ll 4B$. Moreover, in real laser and laser-like (gaseous maser, paramagnetic maser, phaser, NMR-laser e.a.) systems it is difficult (if not possible at all) to rich $A$ more than $O(10)$.

Introducing

$$\left.\begin{array}{l} \tau_\varepsilon = \dfrac{\tau}{\varepsilon_1} \equiv \widetilde{\omega}_{rel}\tau \equiv \omega_{rel} t; \\ m = \widetilde{\omega}_m \varepsilon_1 \equiv \widetilde{\omega}_m / \widetilde{\omega}_{rel} \equiv \omega_m / \omega_{rel} \end{array}\right\}, \qquad (3.1.34)$$

we have

$$m\tau_\varepsilon = \widetilde{\omega}_m \tau = \omega_m t. \qquad (3.1.35)$$

Now let we fulfil the replacement of dynamical variables:

$$J = (A-1)x; \quad n = \varepsilon_2 y + 1, \qquad (3.1.36)$$

or

$$x = \dfrac{J}{A-1}; \quad y = \dfrac{n-1}{\varepsilon_2}, \qquad (3.1.37)$$

and transform (3.1.30), (3.1.31) to the following form:

$$\dfrac{dx}{d\tau_\varepsilon} = x[y - \delta_\varepsilon \cos(m\tau_\varepsilon)] ;, \qquad (3.1.38)$$

$$\dfrac{dy}{d\tau_\varepsilon} = 1 - \varepsilon_1 y - (1+\varepsilon_2 y)x, \qquad (3.1.39)$$

where

$$\delta_\varepsilon = \dfrac{k_m}{\varepsilon_2}, \qquad (3.1.40)$$

### 3.1.4. Converting equations of motion for a periodically modulated phaser to an autonomous system

For nonautonomous dynamical system (3.1.38) – (3.1.39), as for all differential equations of the type

$$\frac{\mathrm{d}}{\mathrm{d}t}\vec{z} = \vec{\Phi}(\vec{z},t,\vec{c}); \quad \vec{z}(t_0) = \vec{z}_0, \tag{3.1.41}$$

the initial value $t = t_0$ commonly speaking can not be assumed as zero. But such the arbitrary choice of the value $t_0$ (which is convenient for computing) is obviously valid for the autonomous systems, having vector field $\vec{\Phi}$ independent on $t$:

$$\frac{\mathrm{d}}{\mathrm{d}t}\vec{z} = \vec{\Phi}(\vec{z},\vec{c}); \tag{3.1.42}$$

$$\vec{z}(t_0) = \vec{z}_0. \tag{3.1.43}$$

The solution of equations (3.1.41) with the initial values $\vec{z}_0$ at $t = 0$ represents the trajectory $\vec{\phi}_t(\vec{z}_0)$ of the system, and the mapping $\vec{\phi}_t : \mathbf{R}^d \to \mathbf{R}^d$ is its flow. Let us formulate in this terms the algorithm of transformation of nonautonomous system with periodic external force and dimension $d$ to equivalent autonomous system with dimension $d+1$.

By introducing new dynamical variable $\zeta = \omega_m t$ the system (3.1.41) may be written in the next form:

$$\left.\begin{array}{l}\dfrac{\mathrm{d}\vec{z}}{\mathrm{d}t} = \vec{\Phi}\left(\vec{z},\dfrac{\zeta}{\omega_m}\right); \quad \vec{z}(0) = \vec{z}_0; \\ \dfrac{\mathrm{d}\zeta}{\mathrm{d}t} = \omega_m; \quad \zeta(0) = \omega_m t_0\end{array}\right\}, \tag{3.1.44}$$

Because the vector field $\vec{\Phi}$ is periodic with the period $T_m = 2\pi/\omega_m$, the system (3.1.44) is periodic with the period $2\pi$. Therefore, the hyperplanes $\zeta = 0$ and $\zeta = 2\pi$ may be glued one to another. Now the phase space converts from the Euclidian space $\mathbf{R}^{d+1}$ to cylindrical one $\mathbf{R}^d \otimes \mathbf{S}^1$, where $\mathbf{S}^1 := [0, 2\pi)$. In the new space the solution of the autonomous system (3.1.44) is the generalized trajectory:

$$\begin{bmatrix}\vec{z}(t)\\ \zeta(t)\end{bmatrix} = \begin{bmatrix}\vec{\phi}_t(\vec{z}_0;t_0)\\ \omega_m t \bmod 2\pi\end{bmatrix}.$$

To conclude, equations of motion for a periodically-modulated phaser reads:

$$\frac{dx}{d\tau_\varepsilon} = x[y - \delta_\varepsilon \cos z] ; \qquad (3.1.45)$$

$$\frac{dy}{d\tau_\varepsilon} = 1 - \varepsilon_1 y - (1 + \varepsilon_2 y)x ; \qquad (3.1.46)$$

$$\frac{dz}{d\tau_\varepsilon} = m , \qquad (3.1.47)$$

where, in accordance with (3.1.29), (3.1.32), (3.1.34), (3.1.40),

$$\varepsilon_1 = [(A-1)B]^{-1/2} ; \; \varepsilon_2 = [(A-1)/B]^{1/2} ; \; m = \omega_m/\omega_{rel} ; \; \delta_\varepsilon = k_m/\varepsilon_2 , \; \omega_{rel} = 1/\varepsilon_1 T_{1S} .$$

Return to standard variables may be fulfilled by the formulae:

$$\left. \begin{array}{l} J = (A-1)x ; \\ n = [(A-1)/B]^{\frac{1}{2}} y + 1 ; \\ t = \omega_m^{-1} z \end{array} \right\}, \qquad (3.1.48)$$

Despite of its simplicity, the described model gives a satisfactory explanation of experimental data for optical-wavelength paramagnetic phasers based on high-quality acoustic Fabry-Perot resonators (see [18], [19]). In Section 3.2 [17], a numerical integration of these equations will be presented to model phenomena of generalized multistability, deterministic acoustic chaos and crises observed in nonautonomous ruby phasers.